\documentclass[prb,twocolumn,superscriptaddress,noshowpacs]{revtex4}
\usepackage{graphicx}
\usepackage{leftidx}
\usepackage{pdfsync}
\usepackage{subfigure}
\usepackage{amssymb}
\usepackage{amsmath}
\usepackage{epstopdf}
\usepackage{color}
\usepackage{bbm}

\newlength{\onecolfig}
\newlength{\twocolfig}
\newcommand{\ion}[2]{\mbox{$^{#2}$#1$^+$}}
\newcommand{\Yb}[1]{\ion{Yb}{#1}}

\newcommand{\unit}[1]{\,\mbox{#1}}

\newcommand{\kHz}{\unit{kHz}}
\newcommand{\MHz}{\unit{MHz}}

\newcommand{\THz}{\unit{THz}}

\newcommand{\us}{\unit{$\mu$s}}

\newcommand{\ket}[1]{\mbox{$\left| #1 \right>$}}

\begin{document}

\title{Shortcuts to adiabaticity by counterdiabatic driving for trapped-ion displacement in phase space}

\author{Shuoming An}
\affiliation{Center for Quantum Information, Institute for Interdisciplinary Information Sciences, Tsinghua University, Beijing 100084, People's Republic of China}
\author{Dingshun Lv}
\affiliation{Center for Quantum Information, Institute for Interdisciplinary Information Sciences, Tsinghua University, Beijing 100084, People's Republic of China}
\author{Adolfo del Campo}
\affiliation{Department of Physics, University of Massachusetts, Boston, MA 02125, USA}
\author{Kihwan Kim}
\affiliation{Center for Quantum Information, Institute for Interdisciplinary Information Sciences, Tsinghua University, Beijing 100084, People's Republic of China}

\begin{abstract}
\noindent
The application of adiabatic protocols in quantum technologies is severely limited by environmental sources of noise and decoherence.
Shortcuts to adiabaticity by counterdiabatic driving constitute a powerful alternative that speed up time-evolution while mimicking adiabatic dynamics.
Here we present  the first experimental implementation of counterdiabatic driving in a continuous variable system, a shortcut to the adiabatic
transport of a trapped ion in the phase space. The resulting dynamics is equivalent to a ``fast-motion video'' of the adiabatic trajectory.
The robustness of this protocol is shown to surpass that of  competing schemes based on classical local controls and Fourier optimization methods.
Our results demonstrate that shortcuts to adiabaticity  provide a robust speedup of quantum protocols of wide applicability in quantum technologies.\\
\end{abstract}
\maketitle

\noindent Adiabatic processes play an essential role in many aspects of quantum technology \cite{CZ12,DS15}.
Quantum adiabatic simulation  exploits adiabatic dynamics to track ground states of complex Hamiltonians facilitating the study of  quantum many-body phenomena \cite{Farhi01,Aharonov08}.
Schemes for scalable ion-trap quantum computer  resort to the adiabatic  transfer of ions between different trap zones \cite{Kielpinski02,Home09}.
Adiabatic dynamics plays as well a key role in holonomic quantum computation \cite{ZR99} and the design of the geometric phase gate \cite{Duan01} with its inherent robustness.
Adiabatic protocols are also essential in quantum thermodynamics whether studying  quantum fluctuations \cite{An14} or the optimization of  quantum thermal machines \cite{Quan07,Deng13,delcampo14,delcampo16}.
These applications are however limited by the requirement  of slow driving that conflicts with the feebleness of quantum coherence when the system of interest is embedded in an environment.\\
\indent According to the adiabatic theorem, a system prepared in a non-degenerate eigenstate will remain in the instantaneous eigenstate during its time evolution under the requirement of slow driving. By contrast, the breakdown of adiabatic dynamics under fast driving couples different energy modes and induces diabatic transitions. Diabatic excitations can however be tailored using shortcuts to adiabaticity (STA) to mimic adiabatic dynamics. Among the available techniques to engineer STA \cite{Torrontegui13}, counterdiabatic driving (CD), relies on the use of an auxiliary control $\hat{H}_{\rm CD}$ to explicitly suppress transitions between different energy eigenstates and enforce parallel transport \cite{Rice03,Berry09}.\\
\indent The transport can be realized by applying a time dependent force $f(t)$ to a harmonic oscillator of mass $m$ and frequency $\omega$, which is described by
\begin{equation}
\label{Transport}
 \hat{H}_{0}=\hat{p}^{2}/2m+m\omega^{2}\hat{x}^{2}/2+f(t)\hat{x}.
\end{equation}
If we increase the force from zero to $f(t)$ slowly, we can transport the ion over a distance $q(t)=-f(t)/m\omega^{2}$. The excitations during the nonadiabatic transport can be seen in the instantaneous frame through the position-shift transformation $e^{iq(t)\hat{p}}$, where we denote $\hbar\equiv1$ throughout the manuscript. In the instantaneous frame, the time-dependent potential minimum is located at $x=0$ and the state is governed by the Hamiltonian $\hat{p}^{2}/2m+m\omega^{2}\hat{x}^{2}/2+\frac{\dot{f}(t)}{m\omega^{2}}\hat{p}$, where a global phase term has been ignored. The first two terms describe the harmonic motion around the potential minimum. The last term is nonlocal in real space and induces diabatic transitions, vanishing only in the adiabatic limit.
The CD suppresses these non-adiabatic transition without slowing down the dynamics by adding the auxiliary term \cite{Torrontegui13,Deffner14}
\begin{equation}
\label{HCDT}
\hat{H}_{\rm CD}=-\frac{\dot{f}(t)}{m\omega^{2}}\hat{p}.
\end{equation}
Because $\hat{p}$ is invariant under the position-shift transformation, diabatic transitions are completely suppressed in the instantaneous reference frame under arbitrarily fast transport.\\
\indent Here we experimentally realize the CD protocol for the nonadiabatic control of a single \Yb{171} ion \cite{Leibfried03,Olmschenk07} as it is transported in phase space. We use a pair of Raman beams to apply the force on the ion and achieve a precise and flexible control of the quantum evolution that allows us to unveil the superior performance of STA based on CD over alternative schemes. Our experiment provides a faithful realization of various STA protocols and is therefore complementary to previous studies on ion transport with time dependent electric fields \cite{Hensinger06,Walther12,Bowler12}.\\

\begin{figure}[!t]
\begin{center}
\includegraphics[width=0.94\onecolfig]{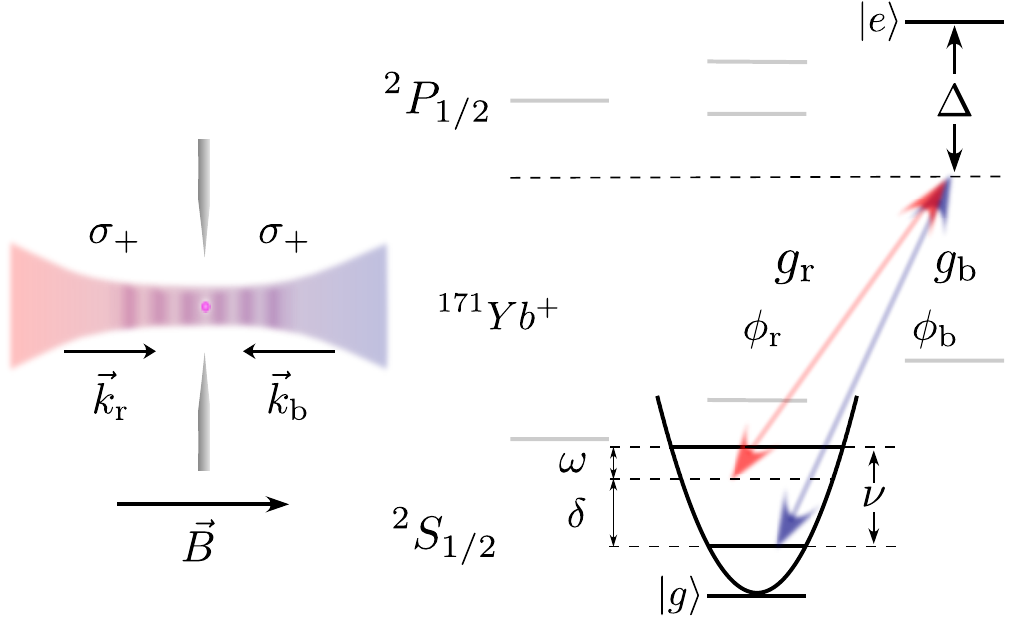}
\end{center}
\caption{\textbf{The scheme of the dragged harmonic oscillator model realized with a trapped \Yb{171}} (energy levels not to scale).
A pair of circularly polarized Raman laser beams with beat note $\delta$ counter propagate along the direction of the transversal motion and the magnetic field $\textbf{B}$. They are far ($\Delta\approx 2\pi\times14\THz$) detuned from the exited state $|e\rangle$, which thus can be adiabatically eliminated. The moving standing wave formed by the lasers shakes the ion with the frequency $\delta$, which is smaller than the transversal trap frequency $\nu=2\pi\times3.1\MHz$ by $\omega=2\pi\times 20\kHz$. In the rotational framework about the beat note frequency, the ion is dragged by the laser-induced force with an equivalent trap frequency $\omega$. By varying the intensity and the phase of the Raman beams, we can control the direction and strength of the displacement of the ion in the phase space.
}
\label{EL}
\end{figure}

\textbf{Results}\\
\textbf{Physical model and quantum control.} In the interaction picture with respect to the harmonic oscillation, the force induced by the lasers as configured in Fig.\ \ref{EL}  is described by
\begin{equation}
  \label{Heff}
\hat{H}_{\rm eff}=f(t) x_{0}\left(\hat{a}e^{-i(\omega t+\phi)}+\hat{a}^{\dagger}e^{i(\omega t+\phi)}\right),
\end{equation}
where $x_{0}=\sqrt{1/2m\omega}$, $f(t) = \Omega(t)\Delta k/2$, $\Omega(t)$ is proportional to the intensity of both Raman beams, $\Delta k$ is the projection of the wave-vectors difference of the Raman beams on the motional axis of the ion and $\phi$ is the phase difference between those two laser beams. Both laser beams are red detuned from the transition between the ground state $\ket{g}=\ket{F=0,m_{F}=0}\ (\leftidx{^2}{S}{_{1/2}})$ and the excited state $\ket{e}=\ket{F=1,m_{F}=1}\, (\leftidx{^2}{P}{_{1/2}})$. Due to the large detuning $\Delta\approx 2\pi\times14\THz$, the excited state $\ket{e}$ is adiabatically eliminated. The effective trap frequency $\omega=2\pi \times 20$ kHz in the interaction frame comes from the difference between the beat-note frequency of the laser beams $\delta$ and the real trap frequency $\nu=2\pi\times3.1\MHz$. The effective mass is given by $m=\frac{\nu}{\omega}M_{\rm Yb}$ ($M_{\rm Yb}$: mass of \Yb{171}). When the phase $\phi=0$, the Hamiltonian (\ref{Heff}) describes a dragged harmonic oscillator, with the dragging term $f(t)\hat{x}$. We can implement the CD term  $h(t)\hat{p}$, where $h(t)=-\frac{\dot{f}(t)}{m\omega^{2}}$, by setting $\phi=-\pi/2$. \\

\begin{figure}[!t]
\begin{center}
\includegraphics[width=0.85\onecolfig]{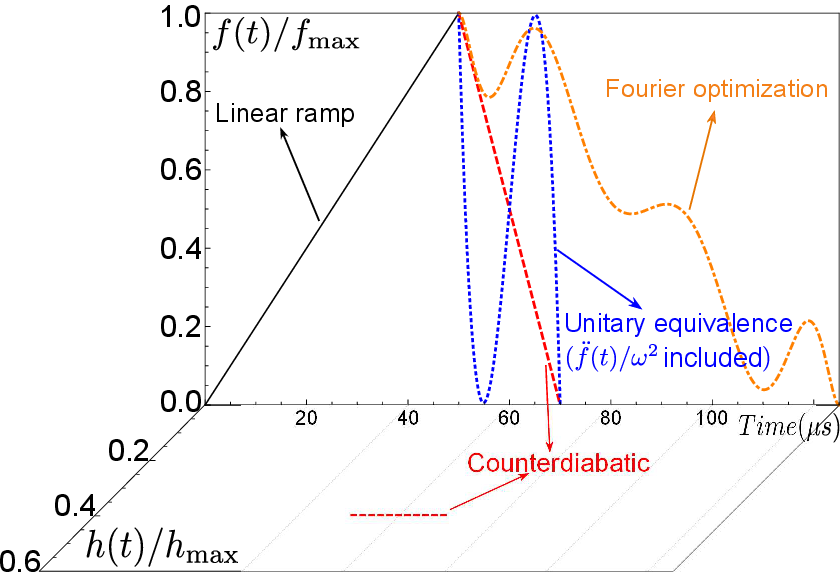}
\end{center}
\caption{\textbf{Time-dependent control fields for different STAs.} In order to measure the phonon excitation after a shortcut, we first use a linear ramp within one period of the harmonic oscillation $2\pi/\omega=50\us$. Then we apply different STA protocols to bring the ion back to its original location. The force $f(t)$ is increased or decreased by changing the intensity of Raman laser beams with $\phi=0$. The function $h(t)$ represents the strength of the CD term proportional to the momentum, which is implemented by applying the laser beams with $\phi=-\pi/2$ during the backward transport. The $f_{\rm max}$ and $h_{\rm max}$ are the maximum values allowed by the common maximum intensity of the laser beams. The smallest shortcut ratio is limited by the maximum laser intensity and we choose the value $s=0.4$ for the CD and UE transport and $s=1.5$ for the Fourier optimization scheme of degree $N=3$.}
\label{Wav}
\end{figure}

\textbf{Counterdiabatic transport.} In the experiment, after Doppler and motional sideband cooling, $\ket{g}$ is prepared with $0.02\pm 0.02$ average phonon number. Because we can not measure the phonon distribution in the interaction picture directly, the STA performance  is probed with a quench echo method \cite{QZ10} in which the ion is first transported adiabatically and then brought back to the initial location using the STA protocol. During the first adiabatic process, we linearly increase the force $f(t)$ from $0$ to  $f_{\rm max}=\Omega_{\rm max}\Delta k/2$ within one period of the harmonic motion $T_{\rm 0}=2\pi/\omega=50\us$, where $\Omega_{\rm max}=2\pi\times378\kHz$ corresponds to the maximum value allowed by the laser. This linear ramp has been well studied in experiments \cite{Couvert08,Bowler12,An14} and can be regarded as perfectly adiabatic. Following it, the force is linearly reduced from $f_{\rm max}$ to $0$ within a duration of $sT_{0}$, where $s$ is defined as the shortcut ratio. The backward dynamics is assisted by turning on the laser to implement the CD term $h(t)\hat{p}$ according to equation\ (\ref{HCDT}). The relation between the strength of the CD term and the shortcut ratio is given by $h(t)\equiv h_{\rm max}/(2\pi s)$. Finally, we apply blue sideband transitions to measure the phonon distribution \cite{Leibfried03}. The time-dependent laser intensity profiles (waveforms) during the forward and backward transport stages are showed in Fig.\ \ref{Wav}. We vary $s$ from $0.95$ to $0.15$ with a step of 0.1 and obtain the final average phonon number $0.016\pm 0.018$, which confirms that the CD protocol does not excite the motion after the transport for any duration. \\
\begin{figure*}[!t]
\centering
    \subfigure
    {\includegraphics[width=0.452\textwidth]{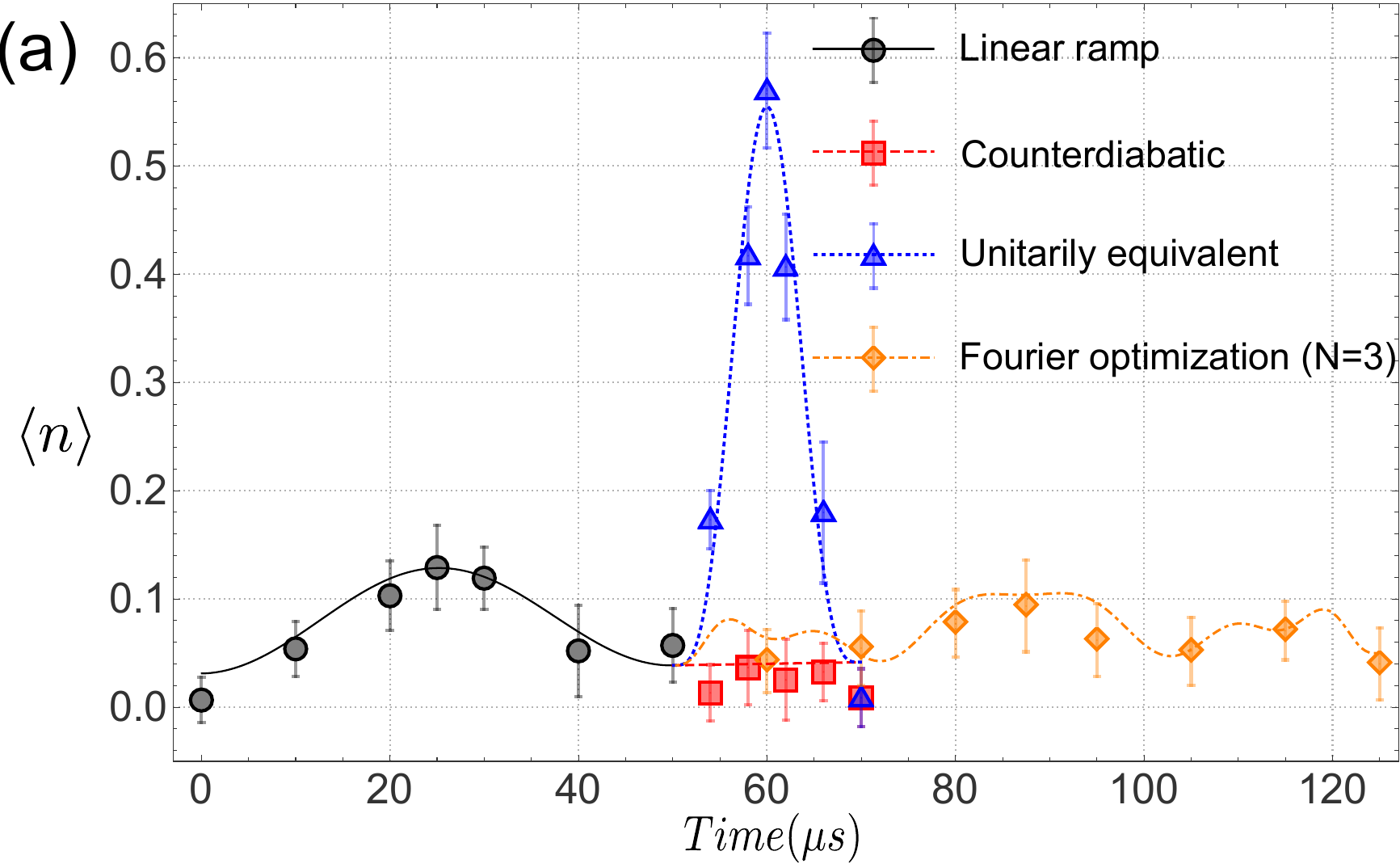}
    \label{Ins}}
    \ \ \ \ \ \ \ \ \ \ \ \
    \subfigure
    {\includegraphics[width=0.45\textwidth]{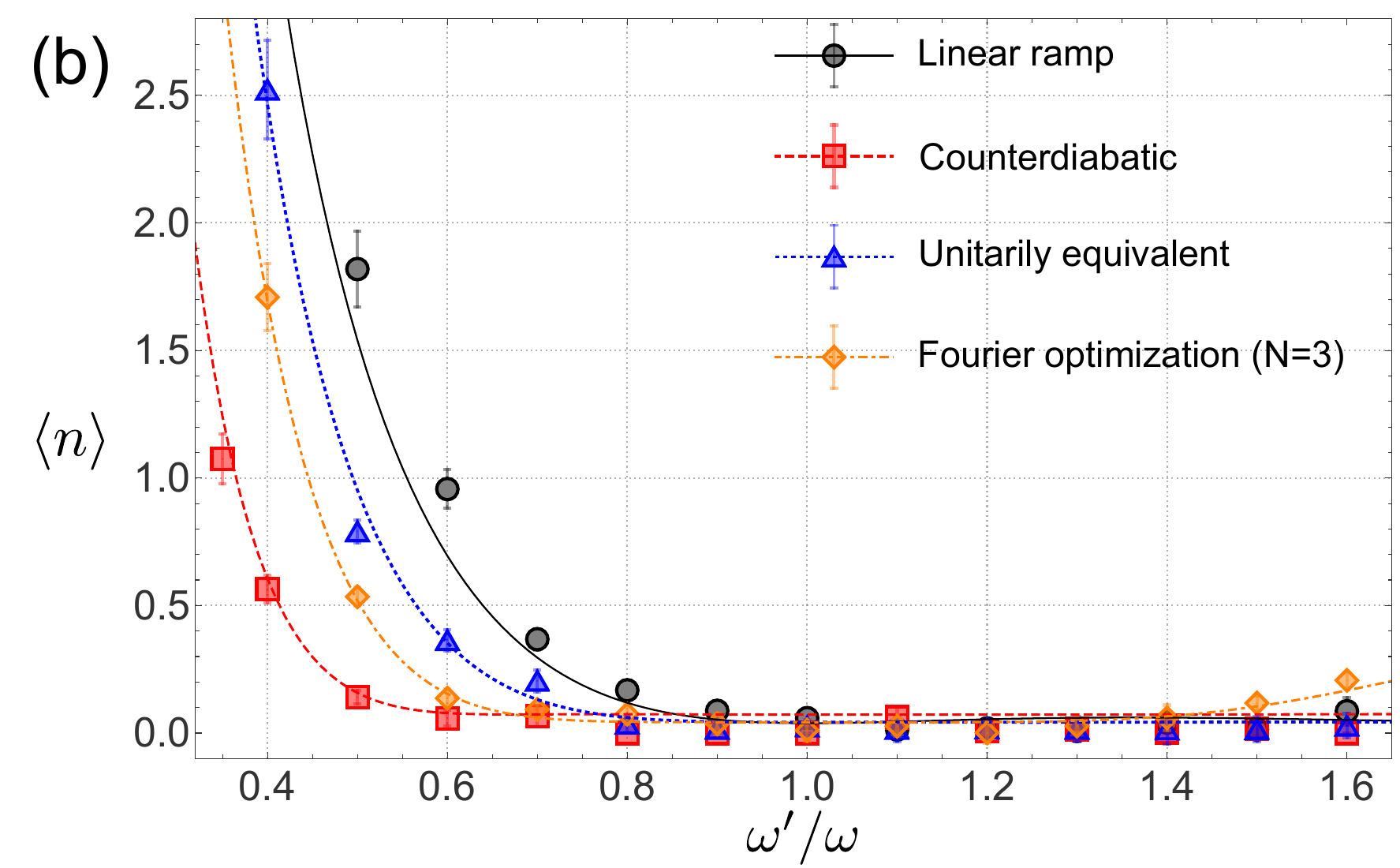}
    \label{Err}}
\caption{\textbf{Phonon excitations in the instantaneous frame during different STA protocols and the robustness against trap frequency errors.}
(a) To measure nonadiabatic excitations during each STA, the shortcut waveform stops at a specific time and the system is transported back adiabatically to the initial position $q=0$. This process brings phonon distributions back to the lab framework for the measurement. Note that only the CD realizes the adiabatic following. (b) To study the robustness of different STA protocols against the trap frequency drift, we change the trap frequency $\omega$ to $\omega'$ during the shortcut transports, whose waveforms are still designed for the nominal trap frequency $\omega=2\pi\times 20\kHz$. Finally we measure the average values of excited phonons. For a fair comparison, we set $s=1.5$ for all three STA protocols. We also test the linear ramp method as reference.
Finally the CD driving is found to be the most robust.
The lines in both figures correspond to the numerical solution of the Lindblad master equation for the noise-average dynamics.
The error bars represent the standard deviation of 200 measurements of the average phonon excitation.
}
\label{InsErr}
\end{figure*}
\indent We also measure phonon excitations in the instantaneous basis during the transport in order to certify that the dynamics is following the adiabatic  ground state. During the forward linear ramp and the backward CD transport, we stop at different instants and add another CD transport with $s=0.15$ to adiabatically change back to the lab frame. As shown in Fig.\ \ref{Ins}, we do not observe any significant excitation during the transport, which confirms that the CD is speeding up the adiabatic trajectory associated with $\hat{H}_0$ as in a ``fast motion video''. We also measure the excitation in the lab framework.\\
\indent Furthermore, the CD is shown to be robust against the trap frequency drift error. We design STA's waveforms with the nominal trap frequency $\omega=2\pi\times 20\kHz$. In the first linear adiabatic ramp, we keep the trap frequency $\omega$, but change the effective trap frequency to $\omega'$ during the STA transport. Then we measure the final average phonon excitation as a function of $\omega'/\omega$. The result in Fig.\ \ref{Err} shows that the CD is extremely robust against the drift of the trap frequency. This feature can be qualitatively explained by the results shown in Fig.\ \ref{Ins}, where almost no excitation appears during the CD driving. Since the higher excited states are more fragile to errors, the protocol with the smaller excitations during the transport is surely more robust. The higher robustness also results from the lower amplitude of the required control field.  In the experiment, for the shortcut ratio $s$ = 0.4, the CD protocol uses three times less intensity than the other protocols, which naturally reduces the amount of noise proportionally.\\
\ \\
\textbf{Unitarily equivalent transport.} The CD stands out among STA protocols for its robustness and the adiabatic following during the
whole evolution. Yet, the realization of the auxiliary control $h(t)\hat{p}$ is hardly feasible with classical electrical fields.
Many efforts have been devoted to identify alternative controls requiring only local potentials \cite{Morsch12,Torrontegui13, Deffner14, Furst14,Odelin14}.
To this end, we resort to controls related to  CD via its unitary equivalence (UE) \cite{Morsch12,Zhang13,Torrontegui13,Deffner14}.
The exact solution to time-dependent Schr\"{o}dinger equation with Hamiltonian $\hat{H}=\hat{H}_{0}+\hat{H}_{\rm CD}$ is given by the adiabatic approximation $\ket{\psi(t)}$ to the dynamics generated by $\hat{H}_{0}$.
Under a momentum-shift transformation $\hat{U}(t)=\exp[-i\dot{f}(t)\hat{x}/\omega^{2}]$, the time evolving state becomes  $\ket{\varphi(t)}=\hat{U}(t)\ket{\psi(t)}$, which is governed by the Hamiltonian,
\begin{equation}
  \label{HA}
  \hat{H}_{U}=\hat{U}\hat{H}\hat{U}^{\dagger}+i\dot{\hat{U}}\hat{U}^{\dagger}=\hat{H}_{0}+\frac{\ddot{f}(t)}{\omega^2}\hat{x},
\end{equation}
where a global phase term has been gauged away. The auxiliary control in the driving Hamiltonian $ \hat{H}_{U}$ can be realized with a local potential.
As long as $\hat{U}(0)=\hat{U}(t_{f})=\mathbb{I}$, the state $\ket{\varphi(t_{f})}$ reproduces exactly the desired target state $\ket{\psi(t_{f})}$ upon completion of the STA protocol. This suggests a route to design the UE transport waveform.
The  boundary conditions $f(0)=0$ and $f(t_{f})=f_{\rm max}$ define the transport problem. Vanishing first-order derivatives $\dot{f}(0)=\dot{f}(t_{f})=0$  guarantee that $\hat{U}(0)=\hat{U}(t_{f})=\mathbb{I}$. Considering that generally we do not suddenly turn on or off control fields, we further impose  second-order boundary conditions $\ddot{f}(0)=\ddot{f}(t_{f})=0$.
These constraints are satisfied by a polynomial waveform $f(t)=10(t/t_{f})^{2}-15(t/t_{f})^{3}+6(t/t_{f})^{4}$ \cite{Torrontegui11, Deffner14}.
In the experiment, we apply the UE transport in the backward process as shown in Fig.\ \ref{Wav}.\\
\indent For the UE transport, we measure the final average phonon number $0.026\pm0.019$ for various shortcut ratios $s$ from $1$ to $0.4$ with a step of $0.1$. As shown in Fig.\ \ref{Ins}, we also examine the process of the UE transport in the instantaneous basis and in the lab framework. We observe large excitations in the process, which shows the UE protocol does not follow the adiabatic evolution, but succeeds in preparing the adiabatic target state at the final stage. As shown in Fig.\ \ref{Err}, the robustness against the drift of the trap frequency is below that of the CD transport. Note that the $f(t)$ used is not the only solution.  Simulation results show that the waveform will be more sensitive to the trap frequency error when higher order boundary conditions are considered. The first-order polynomial waveform can also be used to mimnimize the DC Stark shift during the transport with the electric fields \cite{Lau11}.\\
\ \\
\textbf{Fourier optimization transport.} Finally, we implement the Fourier optimization scheme as proposed in \cite{Odelin14}. When the applied force $f(t)$ for  transport satisfies the conditions  $f(0)=0$ and $\dot{f}(0)=\dot{f}(t_{\rm f})=0$, the final excitation energy can be expressed as the Fourier transform of the acceleration of the force at the trap frequency. In principle, this method allows us to find a driving $f(t)$ that simultaneously minimizes the final excitation energy for an ensemble of  $N$ different trap frequencies. When they are equalized, the final excitation is set by $(\omega'^2-\omega^2)^N$, which enhances the robustness with $N$. The cost of the enhanced robustness is the increase of the amplitude of the control field with the order  $N$. In our experiment, we choose $N=3$ that   results in a oscillatory waveform, shown in Fig.\ \ref{Wav}. The required amplitude of the control field greatly surpasses $f_{\rm max}$ for a small shortcut ratio, thus we only test the scheme for $s=1.5$. The excitation in the instantaneous base and its robustness are shown in Fig.\ \ref{Ins} and Fig.\ \ref{Err}, respectively. \\
\ \\
\textbf{Discussion}\\
We have provided the first realization of shortcuts to adiabaticity based on counterdiabatic driving  in a continuous variable system. By demonstrating the robust adiabatic following, we have shown that the resulting time-evolution  follows a ``fast-motion video'' of the adiabatic dynamics.
This  protocol is also known to be the optimal solution of the quantum brachistochrone problem \cite{Takahashi1302}.
We have further realized  two competing  STA protocols for the transport problem:  local UE driving and Fourier optimization methods.
In the UE scheme,  while the auxiliary control field  takes the form of a  time-dependent linear potential,  its amplitude scales as $\ddot{f}(t)\hat{x}/\omega^{2}\propto s^{-2}$ surpassing the value required for counterdiabatic  driving, $\dot{f}(t)\hat{p}/m\omega^{2}\propto s^{-1}$. We note that by further modulating the trap frequency during  transport, these shortcuts can still be accelerated within a maximum control field with the ``rapid scan method", that has been realized for a two level system \cite{Zhang13}. The total duration can then be  reduced to half for the tested UE protocol reported here. As for the Fourier optimization scheme, its robustness is reduced even with respect to the UE scheme for a given amplitude of the control field, but could be increased with the order $N$ and a higher amplitude of the control field.\\
\indent In our experiment, we demonstrate that the challenging non-local CD term can be generated in the interaction frame. Therefore, our results will be directly influential and beneficial to the other experimental works that require adiabatic evolution in short time and are performed in the interaction picture including quantum thermodynamics, quantum simulation, and quantum computation. The transport of a harmonic oscillator can be a test bath for quantum thermodynamics \cite{An14} or as part of a quantum engine \cite{Deng13,delcampo14,delcampo16}, for which the CD protocol can be used to boost the performance. For many quantum-simulation experiments, adiabatic evolution is essential to prepare a complex ground state of non-trivial Hamiltonian from a simple Hamiltonian whether or not in the interaction frame. The non-trivial ground state of a bosonic Hamiltonian or spin-boson Hamiltonian could be implemented via the CD protocol, overcoming the limitation imposed by the coherence time of the system. The CD protocol can also speed up routines in holonomic quantum computation \cite{ZR99,Duan01,SS15,Zhang14}, and enable the implementation of topological quantum computation with non-Abelian braiding operations \cite{Oppen15} that need not be adiabatic.

\ \\
\textbf{Methods}\\
\textbf{The dragged harmonic oscillator model.}
As mentioned in the main text, the Hamiltonian of the dragged harmonic oscillator in the interaction picture about the harmonic motion is equation (\ref{Heff}).
Here we apply a pair of Raman beams to the ion with a beatnote which is red detuned to the real trap frequency $\nu$ with the nominal trap frequency $\omega$ to simulate this Hamiltonian.
The interaction Hamiltonian is $\frac{\Omega\Delta k}{2}\sqrt{1/(2M_{Yb}\nu)}(\hat{a}e^{-i(\omega t+\phi)}+\hat{a}^{\dagger}e^{i(\omega t+\phi)})$, which equals the equation (\ref{Heff}) when $f(t)=\Omega(t)\Delta k/2$ and $x_{0}=\sqrt{1/(2m\omega)}=\sqrt{1/(2M_{Yb}\nu)}$, where the effective mass $m=\frac{\nu}{\omega}M_{Yb}$.\\
\ \\
\textbf{Dynamics in the instantaneous basis.}
To study the STA dynamics we measure phonon excitations in the instantaneous basis during the transport, and use a short CD to change to the lab frame, where the measurements can be made. To choose the protocol for the frame change, we measure the fidelity of different STA with various shortcut ratios and find that the CD transport with the smallest shortcut ratio $s=0.15$ is optimal. In addition to its robustness against the trap frequency error, its shortest duration protects the motion of the ion from the heating effect.\\

\textbf{Acknowledgments}
We thank Mathieu Beau and Bala Sundaram for a critical reading  of the manuscript. This work was supported by the National Basic Research Program of China under Grants 2011CBA00300 (2011CBA00301), the National Natural Science Foundation of China 11374178, 11574002. Funding support from UMass Boston (project P20150000029279) is acknowledged.

\clearpage
\onecolumngrid
\section*{Supplementary Information}
\section*{A. Dragged harmonic oscillator model}
\indent
In this section we provide a microscopic derivation of the dragged harmonic oscillator model.
We utilize one of the transverse motional modes of a single \Yb{171} ion trapped in the linear four rod Paul trap.
The energy levels and all the definitions of symbols are shown in Fig. 1 of the main text.

The internal state of the ion is $c_{g}(t)\ket{g}+c_{e}(t)\ket{e}$.
We define the slow varying amplitude $C_{i}(t)\equiv c_{i}(t)e^{i\omega_{i}t}\ (i=g,\ e)$, where $\omega_{i}$ is the energy of the corresponding internal state.
Using the dipole approximation we can write the equation of motion
\begin{align}
\label{C}
  i\dot{C}_{g}(t)=&\frac{C_{e}(t)}{2}\left(g_{b}e^{-i(\vec{k}_{b}\cdot\vec{x}+\Delta t)}+g_{r}e^{-i(\vec{k}_{r}\cdot\vec{x}+(\Delta+\delta) t-\phi)}\right),\\
  i\dot{C}_{e}(t)=&\frac{C_{g}(t)}{2}\left(g_{b}^{*}e^{i(\vec{k}_{b}\cdot\vec{x}+\Delta t)}+g_{r}^{*}e^{i(\vec{k}_{r}\cdot\vec{x}+(\Delta+\delta) t-\phi)}\right).
\end{align}
In the rotation wave approximation (RWA), we adiabatically eliminate the excited state $\ket{e}$ and get the equation of motion  for the ground state $\ket{g}$ only
\begin{equation}
  i\dot{C}_{g}(t)=-\Omega C_{g}(t)-\frac{\Omega}{2}\left(e^{i(\Delta\vec{k}\cdot\vec{x}+\delta t-\phi)}+H.c.\right)C_{g}(t).
\end{equation}
In the right hand side of the equation, the first term $-\Omega=-\dfrac{g_{b}g_{r}^{\star}}{2\Delta}$ before $C_{g}$  is the AC Stark shift and $\delta \vec{k}=\vec{k}_{r}-\vec{k}_{b}$ and $\phi=\phi_{r}-\phi_{b}$.
If we neglect the AC Stark shift and only consider one of the motional modes (others are ignored by the RWA) in the Lamb-Dicke regime, we find the effective Hamiltonian for the ground state
\begin{equation}
  \hat{H}_{eff}=\frac{\Omega \Delta k}{2}x_{0}\left(\hat{a}e^{-i(\omega t+\phi)}+H.c.\right).
\end{equation}
Here a global phase of $-\pi/2$ has been added.
In this experiment, $\Omega$, $\phi$ and $\omega$ can be changed to control the amplitude of the force term $f(t)\hat{x}$, realize the momentum term $h(t)\hat{p}$ and vary the nominal trap frequency.
As a result, the dragged harmonic oscillator can be fully controlled.
\begin{figure}[htbp]
\includegraphics[width=0.7\twocolfig]{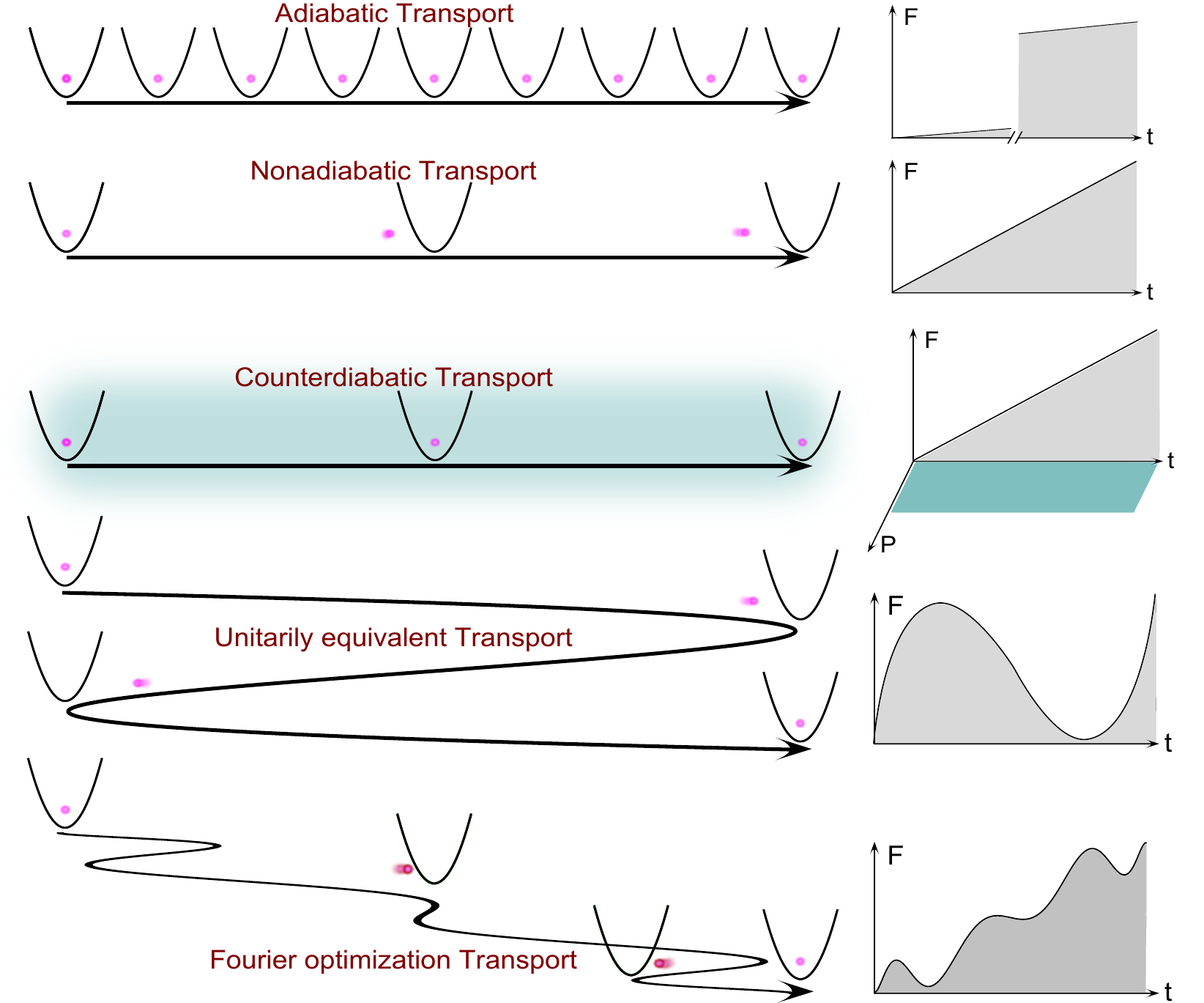}
\caption{Illustration of the different transport protocols experimentally studied. The panels in the right hand side display the time-dependent force applied in the different driving schemes.}
\label{pro}
\end{figure}

\indent In Fig. \ref{pro}, we illustrate different transport protocols that we experimentally study. Sufficiently slow driving corresponds  to the adiabatic transport. Nonadiabatic excitations appear when the protocol is sped up. To suppress them a global interaction according to the CD protocol is applied. As an alternative to this nonlocal interaction, we consider a unitarily equivalent protocol where the dynamics is assisted by a local potential so as to mimic adiabaticity in the preparation of  the final target state.  A variant of this later protocol relies on the the Fourier optimization scheme, whereby the time-dependent force is designed such that the Fourier transform of the acceleration of the displacement vanishes when the angular frequency matches the trap frequency. The resulting protocol involves an oscillatory trajectory,  see Section F for further details.
All these protocols can be realized in the dragged harmonic oscillator model.

\section*{B. Measurement of dynamics in the instantaneous basis }
\begin{figure}[htbp]
\centering
\includegraphics[width=0.5\twocolfig]{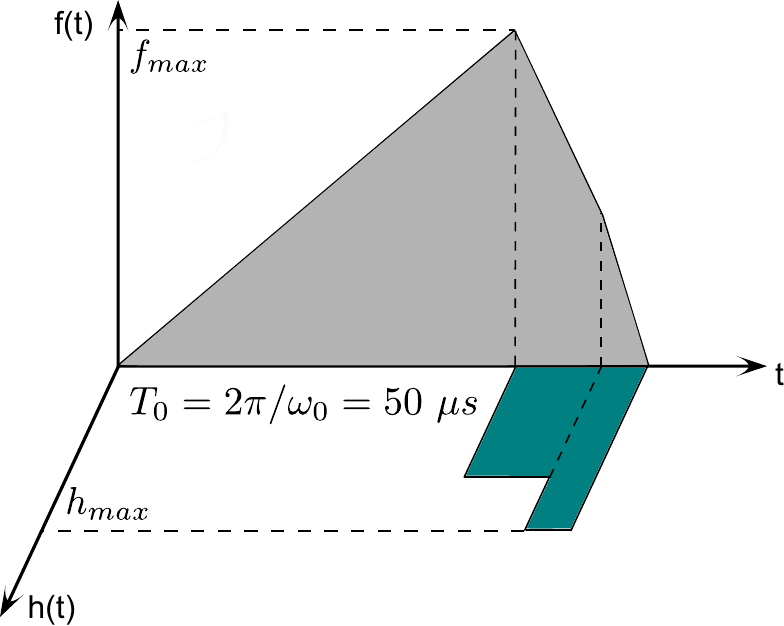}
\caption{In the forward process, we linearly increase the force from zero to $f_{max}$. Then we apply the CD protocol to bring the ion back to the original place. In order to measure the phonon excitation in the instantaneous frame, we apply the additional CD protocol in the end. The final CD transport has the shortcut ratio $s=0.15$. We also apply the same CD protocol to study the dynamics of UE protocol and the Fourier optimization scheme in the instantaneous basis.}
\label{back}
\end{figure}

\indent In order to measure the phonon excitation in the instantaneous basis during the transport, we stop the protocol during the process and adiabatically bring back the phonon system to the lab framework by performing a CD with $s=0.15$ as shown in Fig.\ \ref{back}.

\section*{C. Performance of STA depending on the shortcut ratio}
\begin{figure*}[htbp]
\centering
    \subfigure
    {\includegraphics[width=0.8\onecolfig]{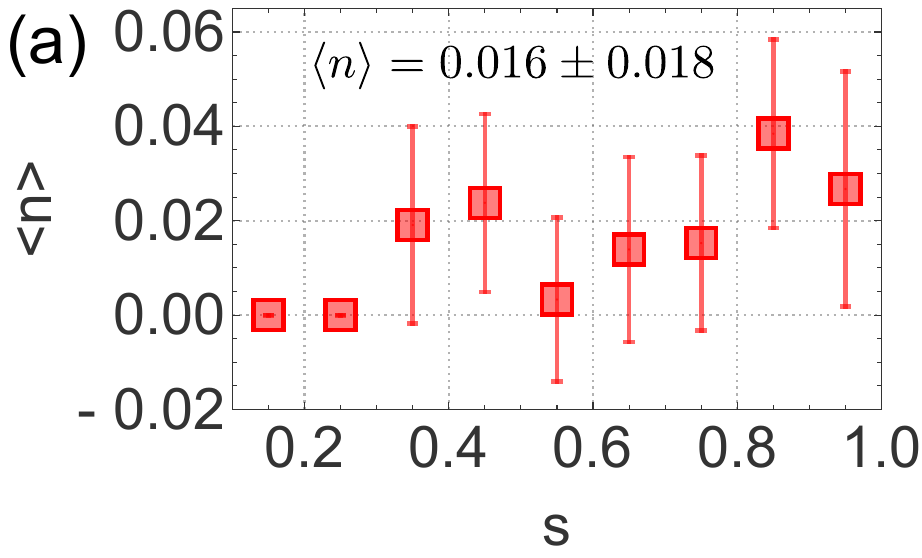}
    \label{cd1}}
    \ \ \ \ \ \ \ \ \ \ \ \
    \subfigure
    {\includegraphics[width=0.8\onecolfig]{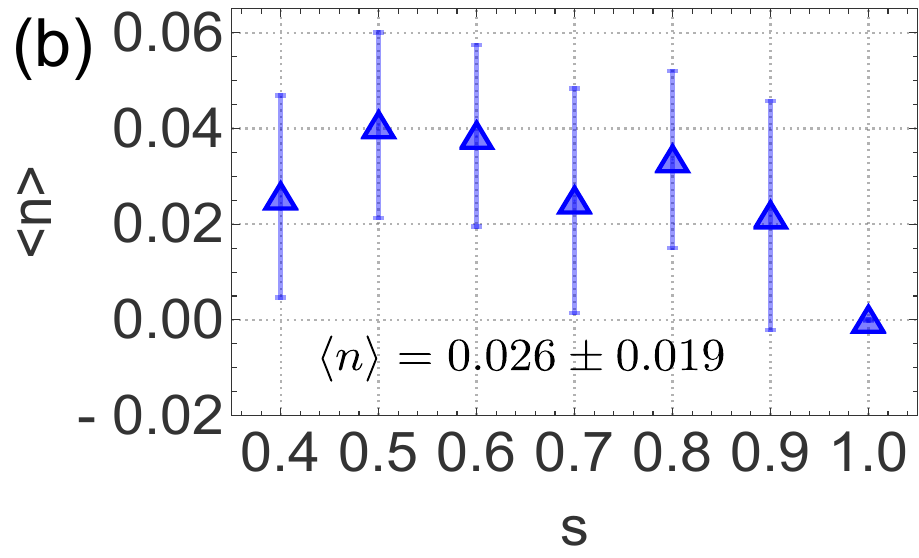}
    \label{ue1}}
\caption{Performance of STAs depending on the shortcut ratio. (a) We apply the CD transport with different shortcut ratios to bring the ion back after the first linear ramp. The final phonon excitations are recorded against different shortcut ratios. We do not observe any degrade of the performance of the CD protocol for the minimum $s=0.15(\approx\frac{1}{2\pi})$ allowed by the maximum laser intensity. (b) We apply the UE transport with different shortcut ratios to bring the ion back after the first linear ramp. The final phonon excitations are recorded against different shortcut ratios. The curves are numerical simulated results and the points are the experiment results. The performance of the UE is good even for the minimum $s=0.4$ allowed by the maximum laser intensity. Note that the smallest shortcut ratio $s$ for the UE is larger than that of the CD protocol and the final excitations are larger than those of the CD protocols, which indicates the CD protocol is more robust that the UE protocol.
}
\label{cdue}
\end{figure*}

\indent To test the STA's performance against the shortcut ratio $s$, we measure the final phonon excitation after the CD backward process with different shortcut ratios as shown in Fig.\ \ref{cd1}. We do not observe any meaningful excitation in phonon number after the CD protocol from $s=0.15$ to $s=1$.

\indent We also test the final excitations against different shortcut ratios from $s=0.4$ to $s=1.0$ for the UE protocol as shown in Fig. \ref{ue1}. For the maximum laser intensity, the minimum shortcut ratio for the UE protocol is 0.4 in our experimental condition.

\section*{D. The maximum excitation of phonons during the transport in the lab frame}
\begin{figure*}[htbp]
\centering
    \subfigure
    {\includegraphics[width=0.75\onecolfig]{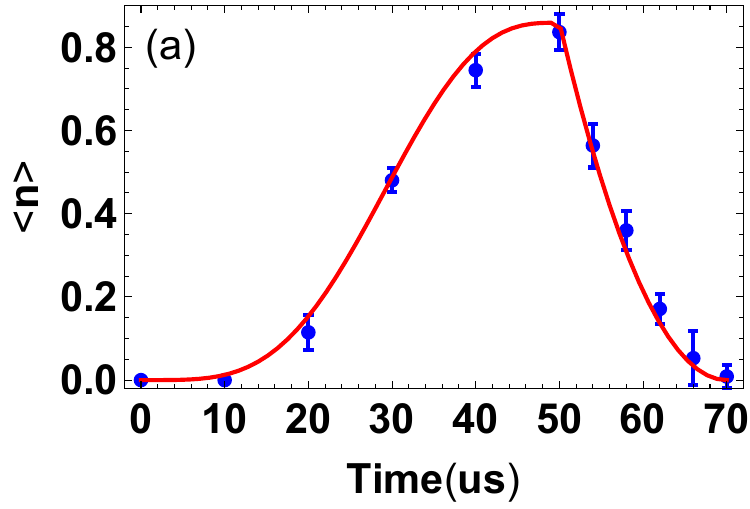}
    \label{IandII}}
    \ \ \ \ \ \ \ \ \ \ \ \
    \subfigure
    {\includegraphics[width=0.75\onecolfig]{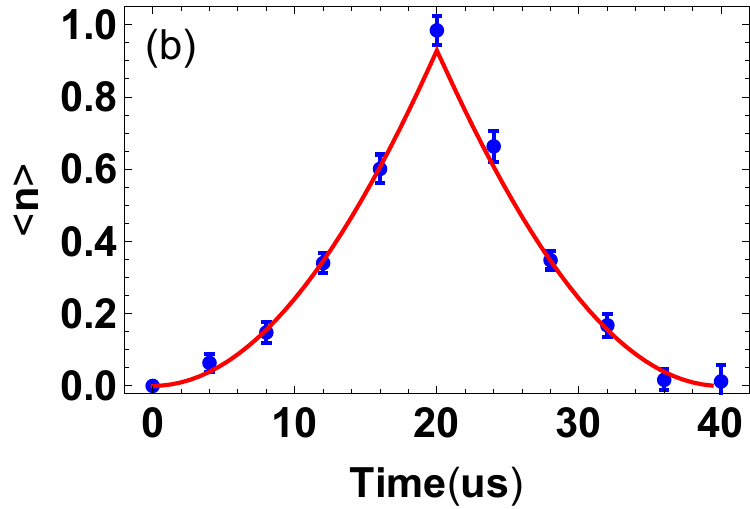}
    \label{IIandII}}
		 \subfigure
    {\includegraphics[width=0.75\onecolfig]{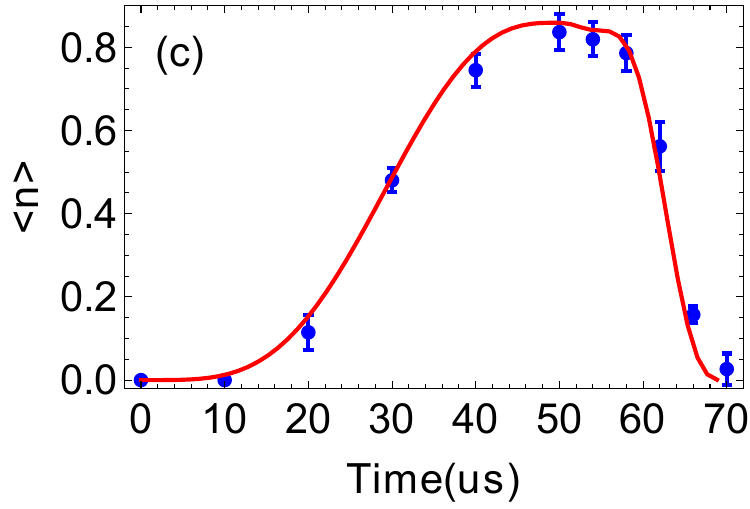}
    \label{IandIII}}
    \ \ \ \ \ \ \ \ \ \ \ \
    \subfigure
    {\includegraphics[width=0.75\onecolfig]{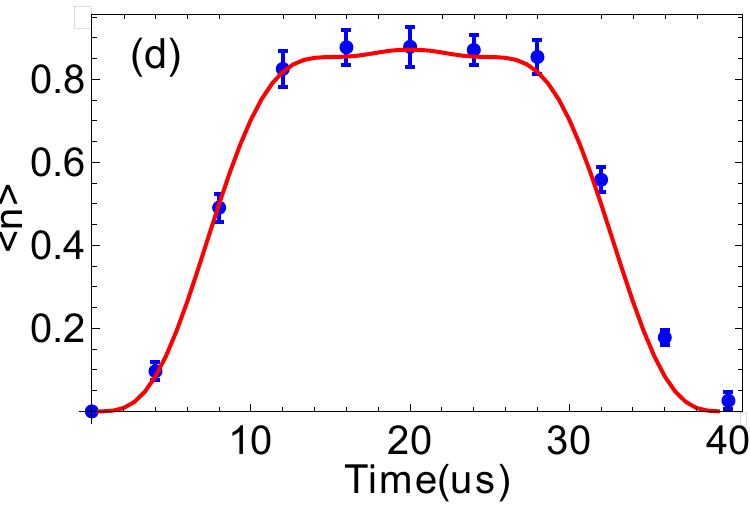}
    \label{IIIandIII}}
		\caption{The excitations in the lab framework for (a,b) counterdiabatic driving and (c,d) the unitarily equivalent driving.
(a) We use the linear ramp to push out and the counterdiabatic transport to bring back.
In the main paper, we stop and add another counterdiabatic transport to bring the population in the instantaneous framework back to the lab framework for the measurement.
Here we stop at a specific time and directly apply blue sideband transitions to measure the excited average phonon in the lab framework.
(b) We use only the counterdiabatic driving for the pushing out and bringing back processes.
We stop at a specific time and measure the excited phonon in the lab framework.
(c) We use the linear ramp to push out and the unitarily equivalent transport to bring back.
In the main paper, we stop and add another counterdiabatic transport to bring the population in the instantaneous framework back to the lab framework for the measurement.
Here we stop at a specific time and directly apply the blue sideband transition to measure the excited average phonon in the lab framework.
Here the shortcut ratio is $0.4$.
(d) We use only the unitarily equivalent driving for the pushing out and bringing back processes.
We stop at a specific time and measure the excited phonon in the lab framework.
}
\label{Tran}
\end{figure*}

Additional data for the tests of the CD and the UE protocol are reported in Fig.\ \ref{Tran}.
Instead of the excitation in the instantaneous framework, we also measure the excited average phonon number in the lab framework.
The data confirm that we actually transport the ion in the lab framework.

\section*{E. The waveforms for different UE transports}
\begin{figure}[htbp]
\includegraphics[width=0.45\twocolfig]{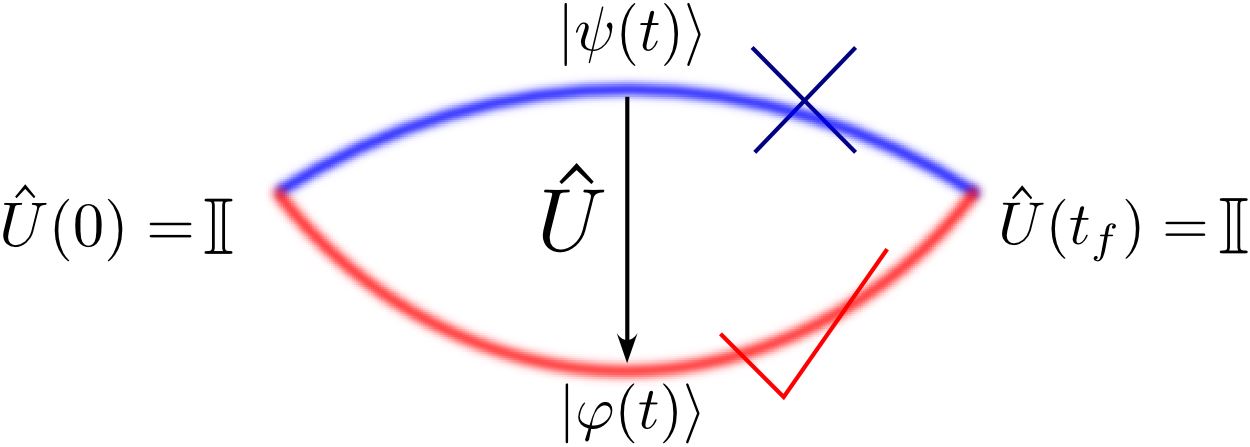}
\caption{The UE protocol.
The transformation $\hat{U}(t)$ is used to find a UE Hamiltonian for the full CD Hamiltonian. In the UE protocol, the STA transport can be assisted exclusively with a local potential. For the modified dynamics to connect the same initial and final states, we impose the zeroth order boundary conditions $\hat{U}(0)=\hat{U}(t_{f})=\mathbbm{I}$ and the first order boundary conditions $\dot{f}(0)=\dot{f}(t_{f})=0$.}
\label{UD}
\end{figure}

\begin{figure}[htbp]
\includegraphics[width=0.75\twocolfig]{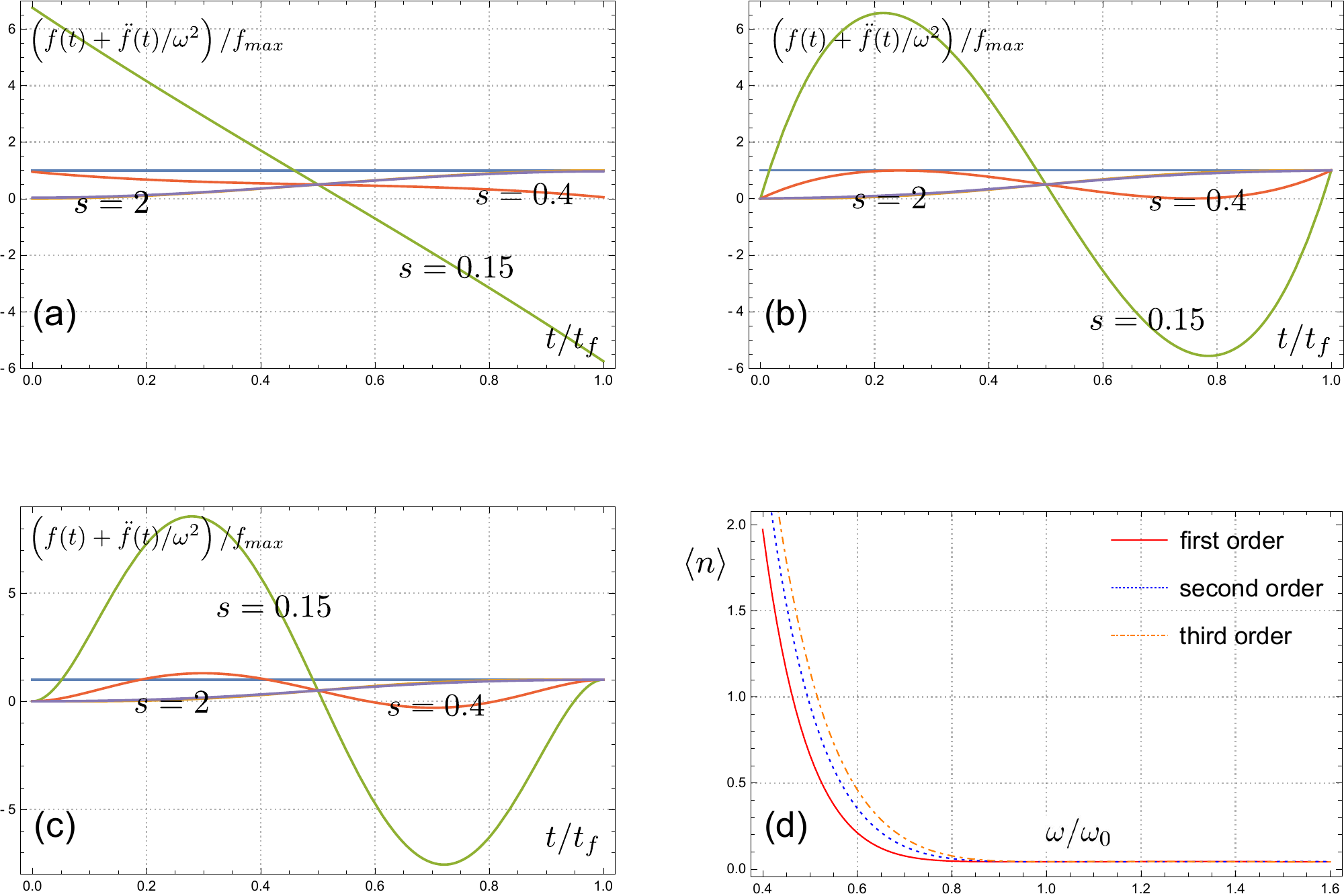}
\caption{Waveforms of (a) the first, (b) second and (c) third order polynomials and (d) their robustness against trap frequency errors. In (d), the robustness for different solutions against the trap frequency errors is compared for the case of $s=1$.
}
\label{XNR}
\end{figure}

Here we provide waveforms for different UE transports. In order to remove the non-local control term proportional to $\hat{p}$ in the CD protocol, a momentum shift transformation $\hat{U}(t)$ is applied. As shown in Fig.\ \ref{UD}, it is necessary to fulfill the zeroth and the first boundary conditions for $f(t)$, which are $\hat{U}(0)=\hat{U}(t_{f})=\mathbbm{I}$ and $\dot{f}(0)=\dot{f}(t_{f})=0$, respectively. However, it is not the sufficient condition. We can impose higher-order boundary conditions for the transports. Fig.\ \ref{XNR}(a)-(c) summarizes the waveforms for different orders of polynomial solutions with different shortcut ratios.

Fig.\ \ref{XNR}(d) shows the comparison of the robustness against the trap frequency errors for the case of $s=1$.
The robustness of the first order waveform is the best, which is consistent to the result in Ref. \cite{Lau11}.
However, it requires the instantaneous switching of the control field, which is difficult in our experimental system. Therefore, we use the second order polynomial waveform in the experiment.\\

\section*{F. Fourier Optimization Scheme}
\indent In this section we briefly describe the STA protocol based on the Fourier optimization scheme, which is well-explained in Ref. \cite{Odelin14} for the self-consistency of the manuscript. The method exploits the fact that when a force $f(t)$ is applied to a particle in the ground state from $f(0)=0$ to $f(t_{f})$ with the first order boundary conditions $\dot{f}(0)=\dot{f}(t_{f})=0$, the final excitation energy is written by
\begin{align}
\label{de}
  \Delta E(\omega_{0})=\left|\int_{0}^{t_{f}}\ddot{f}(t')e^{-i\omega t'}dt'\right|/m\omega^{2},
\end{align}
where $\omega$ is the trap frequency. If we'd like to optimize the protocol for two-different trap frequencies $\omega_{1}, \omega_{21}$, we have to minimize the excitation $\Delta E(\omega_{1})$ and $\Delta E(\omega_{2})$ with proper $f(t)$. This is the principle to design the two-frequency transport waveform. If we set $\omega_{1}=\omega_{2}=\omega$, the robustness of the protocol against the trap frequency error is increased. In Ref. \cite{Odelin14}, an auxiliary function $g(t)$ is introduced to satisfy the following condition,
\begin{align}
  \label{g}
  \ddot{f}(t)=\dfrac{d^{4} g}{dt^{4}}+(\omega^2_{1}+\omega_{2}^{2})\dfrac{d^{2}g}{dt^{2}}+\omega_{1}^{2}\omega_{2}^{2}g(t),
\end{align}
with the boundary conditions $g(0)=g(t_{f})=\dot{g}(0)=\dot{g}(t_{f})=\ddot{g}(0)=\ddot{g}(t_{f})=\dddot{g}(0)=\dddot{g}(t_{f})=0$.
Combining Eqs. (\ref{de}) and (\ref{g}), we get
\begin{align}
  \Delta E(\omega)=\left|(\omega^{2}-\omega^{2}_{1})(\omega^{2}-\omega^{2}_{2})
  \int_{0}^{t_{f}}e^{-i\omega t'}g(t')dt'\right|
\end{align}
and find $\Delta E(\omega_{1})= \Delta E(\omega_{2})=0$. Therefore, we can choose a polynomial function $g(t)$ to satisfy the boundary conditions for $g(t)$ as well as $f(t)$. For this kind of the order $N=2$, we can choose $g(t)\propto (t/t_{f})^{4}(1-t/t_{f})^{4}(1-2t/t_{f})$. The technique can be generalized for the multiple frequency cases by satisfying the condition of $\Delta E(\omega_{i})=0$ for a set of frequencies $\{\omega_i|i=1,\dots,N\}$. By setting these frequencies to be degenerate, the protocol exhibits an enhanced robustness against frequency errors with the order $N$.
For the $N$-order case $g(t)\propto (t/t_{f})^{2N}(1-t/t_{f})^{2N}(1-2t/t_{f})$, from which one can
derive $f(t)$. In the main paper we show the waveform of order $N=3$ for $s=1.5$ due to the limitation of $f(t)/f_{max}<1$.
Here we compare waveforms of different values of $N$ and their robustness against the trap frequency errors.
We can see the maximum amplitude of the control field of this protocol is extremely sensitive to $N$ and $s$.
And for a fixed protocol duration a higher robustness can be achieved by increasing the amplitude of the control field.

\begin{figure}[htbp]
\includegraphics[width=0.75\twocolfig]{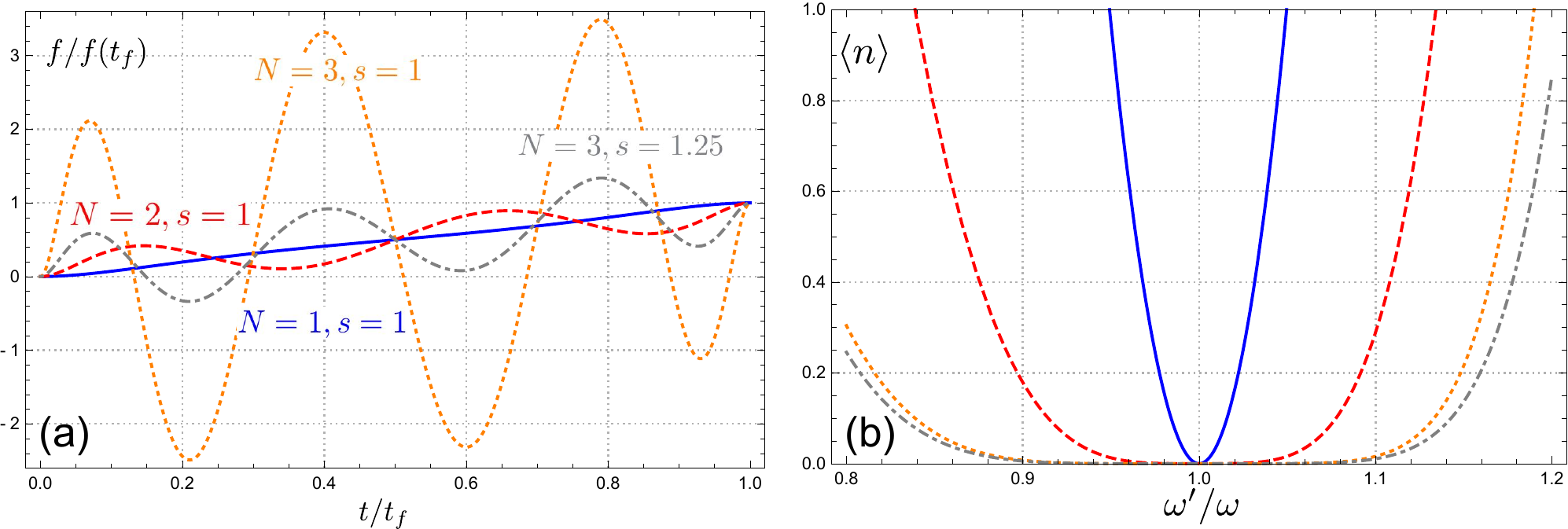}
\caption{ Waveforms for the Fourier optimization scheme of different order $N$ and their robustness against the trap frequency errors.
(a) The maximum amplitude of $f(t)$ increases rapidly with the order $N$ and decreases quickly with the shortcut ratio $s$.
(b) With the same $s=1$, the robustness would be better for a larger order $N$. With the same order $N=3$, the robustness would be better for a larger shortcut ratio $s$.
}
\label{DM}
\end{figure}


\end{document}